\documentclass[twocolumn, aps, prd, showpacs, nofootinbib, superscriptaddress]{revtex4}

\usepackage{graphicx}
\usepackage[colorlinks=true, linkcolor=blue, anchorcolor=blue, citecolor=blue, urlcolor=blue]{hyperref}
\usepackage{natbib}
\usepackage{amsmath}
\usepackage{txfonts}

\newcommand{\diff}{\mathrm{d}}
\newcommand{\scsc}[1]{{\scriptscriptstyle{#1}}}
\newcommand*\Bell{\ensuremath{\boldsymbol\ell}}

\begin{document}

\title{Cosmological parameter constraints from CMB lensing with cosmic voids}
\author{Teeraparb Chantavat}\thanks{E-mail: teeraparbc@nu.ac.th}
\affiliation{ThEP's Laboratory of Cosmology and Gravity, The Institute for Fundamental Study \\ ``The  Tah Poe Academia Institute", Naresuan University, Phitsanulok, 65000, Thailand}
\affiliation{Thailand Center of Excellence in Physics, Ministry of Education, Bangkok, 10400, Thailand}
\author{Utane Sawangwit}
\affiliation{National Astronomical Research Institute of Thailand (NARIT), Chiang Mai, 50200, Thailand}
\author{P. M. Sutter}
\affiliation{UPMC Univ Paris 06, UMR7095, Institut d'Astrophysique de Paris, F-75014, Paris, France}
\affiliation{CNRS, UMR7095, Institut d'Astrophysique de Paris, F-75014, Paris, France}
\affiliation{INFN - National Institute for Nuclear Physics, via Valerio 2, I-34127, Trieste, Italy}
\affiliation{INAF - Osservatorio Astronomico di Trieste, via Tiepolo 11, I-34143, Trieste, Italy}
\author{Benjamin D. Wandelt}
\affiliation{UPMC Univ Paris 06, UMR7095, Institut d'Astrophysique de Paris, F-75014, Paris, France}
\affiliation{CNRS, UMR7095, Institut d'Astrophysique de Paris, F-75014, Paris, France}
\affiliation{Departments of Physics and Astronomy, University of Illinois at Urbana-Champaign, Urbana, IL 61801, USA}


\date{\today}

\begin{abstract}
We investigate the potential of using cosmic voids as a probe to constrain cosmological parameters through the gravitational lensing effect of the cosmic microwave background (CMB) and make predictions for the next generation surveys.  By assuming the detection of a series of $\approx 5 - 10$ voids along a line of sight within a square-degree patch of the sky, we found that they can be used to break the degeneracy direction of some of the cosmological parameter constraints (for example $\omega_b$ and $\Omega_\Lambda$) in comparison with the constraints from random CMB skies with the same size area for a survey with extensive integration time.  This analysis is based on our current knowledge of the average void profile and analytical estimates of the void number function.  We also provide combined cosmological parameter constraints between a sky patch where series of voids are detected and a patch without voids (a randomly selected patch).  The full potential of this technique relies on an accurate determination of the void profile to $\approx 10$\% level.  For a small-area CMB observation with extensive integration time and a high signal-to-noise ratio, CMB lensing with such series of voids will provide a complementary route to cosmological parameter constraints to the CMB observations.  Example of parameter constraints with a series of five voids on a $1.0^{\circ} \times 1.0^{\circ}$ patch of the sky are $100\omega_b = 2.20 \pm 0.27$, $\omega_c = 0.120 \pm 0.022$, $\Omega_\Lambda = 0.682 \pm 0.078$, $\Delta_{\mathcal{R}}^2 = \left(2.22 \pm 7.79\right) \times 10^{-9}$, $n_s = 0.962 \pm 0.097$ and $\tau = 0.925 \pm 1.747$ at 68\% C.L.
\end{abstract}

\maketitle

\section{Introduction}
\label{sec:introduction}

Observations of the cosmic microwave background (CMB) of the Universe have provided a wealth of information about the initial conditions and the structure of our early Universe (for a recent review see Ref.~\citep{Aghanim_ea2008}).  Recent observations of the CMB \citep{Hinshaw_ea2013, PlanckXVI} have shown that our Universe is highly Gaussian with a nearly scale-invariant power spectrum.  This has provided our picture of the Universe as the standard model called the inflationary $\Lambda$CDM model \citep{Carroll_Press1992}.

In the $\Lambda$CDM model, the Universe is homogeneous and isotropic on large scales. However, on small scales, the hierarchical clustering of matter leads to formations of complex cosmic structure such as clusters of galaxies, walls, filaments and voids \citep{Boylan-Kolchin_ea2009}.  Among these objects, voids occupy a vast majority of space and hence provide the largest volume-based test on theories of structure formation \citep{Biswas_ea2010, Bos_ea2012}.  Recently cosmic voids are being continually found, amounting to releases of public void catalogs \citep{Pan_ea2012, Sutter_ea2012a, Sutter_ea2014a}. 

The CMB signal from the surface of last scattering has traversed the Universe for 13.8 billion years to reach us, passing through intervening clusters and voids along the line of sight.  The trajectories of CMB photons are bent toward gravitating matter due to the distortion of spacetime caused by gravitational lensing \citep{Blanchard_Schneider1987}.  The gravitational lensing sources distort the CMB temperatures, giving rise to the transfer of CMB an angular power spectrum to smaller scales \citep{Smith_ea2006}.  The secondary anisotropies due to lensing effects add cosmological information on the growth of the structure and local curvature of the Universe.  The scenario is reversed when voids are acting as the sources of gravitational lenses.  The delensing effect of voids has been investigated and recently observed through the distortions of background galaxies by a stacking method which enhances the signal \citep{Clampitt_Jain2015, Higuchi_ea2013, Krause_ea2013, Melchior_ea2014}.  The statistically significant detection of a correlation between voids and the integrated Sachs-Wolfe effect by voids has also been investigated \citep{Cai_ea2014, Hotchkiss_ea2015, Ilic_ea2013, PlanckXIX}.   A precision cosmology with a void is also attainable---the Alcock-Paczy\'{n}ski test could be applied to the morphology of stacked void in order to infer the underlying cosmology with good precision \citep{Lavaux_Wandelt2012, Sutter_ea2012b, Sutter_ea2014b}.

\begin{figure*}\centering
	\includegraphics[scale=0.55]{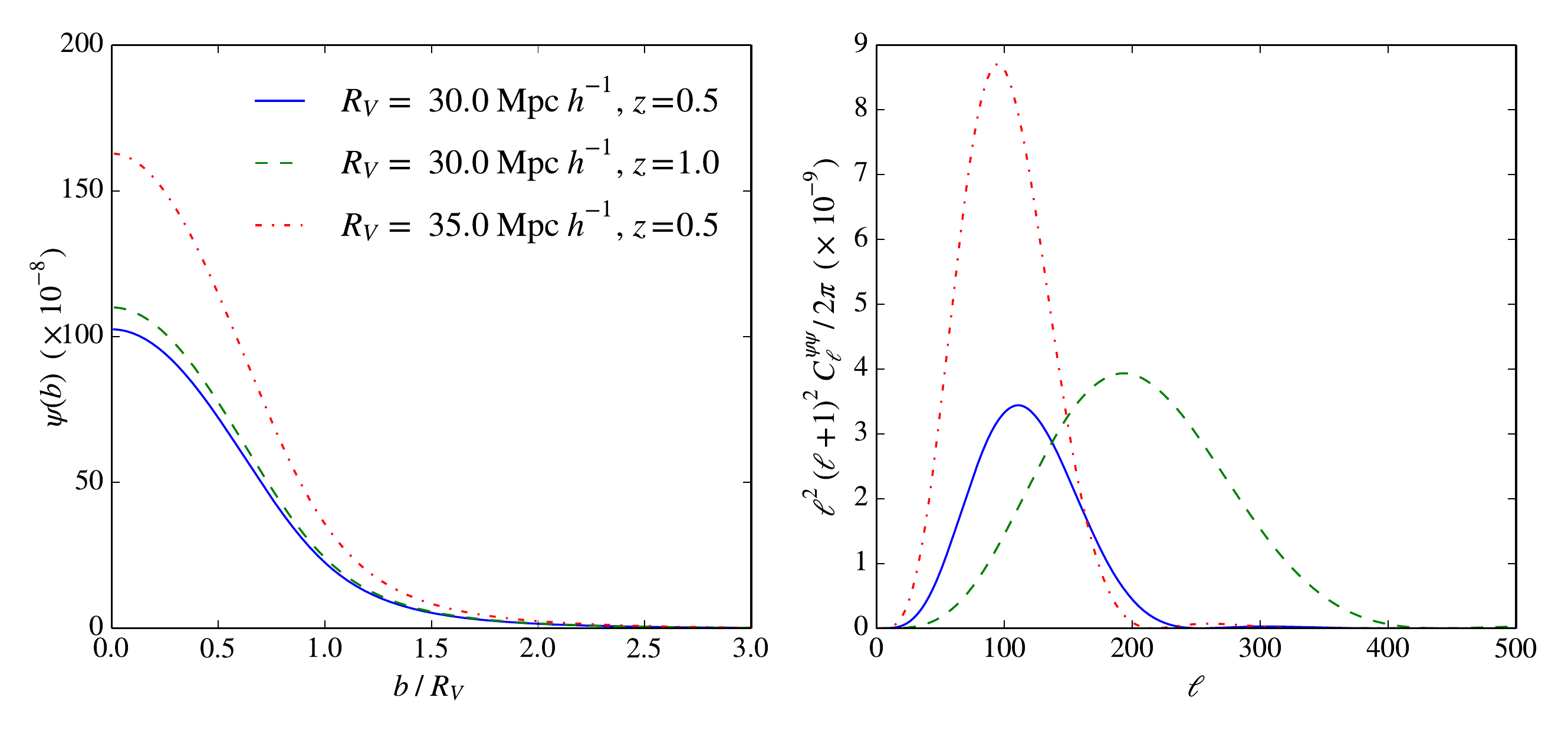}
	\caption{\label{fig1} The lensing potentials of a single void in real space as a function of impact parameter $b$ (left) and their corresponding angular power spectra (right) for voids with $R_V = 30.0\ \mbox{Mpc}\ h^{\scsc{-1}}$ at $z = 0.5$ (solid), $R_V = 30.0\ \mbox{Mpc}\ h^{\scsc{-1}}$ at $z = 1.0$ (dashed) and $R_V = 35.0\ \mbox{Mpc}\ h^{\scsc{-1}}$ at $z = 0.5$ (dot-dashed).}
\end{figure*}

The gravitational lensing effect by voids has a benefit due to the fact that voids have high chance of alignment along a line of sight.  The lensing effect will also be enhanced by having multiple lensing agents (i.e. voids in this case) on the same line of sight.  In addition, the universality of the void profile \citep{Hamaus_ea2014a} could be exploited to predict the lensing effect of voids at a given redshift.  The sensitivity of voids lensing with the cosmological parameters is mainly due to the determination of the comoving angular diameter distance to voids and the linear growth factor.

The goal of this article is to investigate the potential of utilizing voids as probes of cosmology by observing the lensing effect of the CMB.  Our method is based on a comparison with the CMB parameter constraints from a random patch of the sky and a square-degree patch of the sky where a series of voids is detected from large-scale structure surveys.  Throughout this article, our fiducial cosmological parameters for Fisher analysis are $\{100\omega_b, \omega_c, \Omega_\Lambda, \Delta_{\mathcal{R}}^2, n_s, \tau \} = \{2.20, 0.120, 0.682, 2.22 \times 10^{-9},\ 0.962, \ 0.0925\}$, which is consistent with PLANCK + WMAP polarization maximum likelihood cosmological parameters \citep{PlanckXVI} with $w = -1$ and $\Omega_k = 0$ as the standard flat $\Lambda$CDM cosmology.  The matter power spectrum and the angular power spectrum were computed using CAMB\footnote{http://camb.info} \citep{Lewis_ea2000}.

\section{Theory}
\label{sec:theory}

The formalism for CMB lensing correlations, covariance and Fisher information matrices is given in the context of the flat-sky approximation which is appropriate for small-scale CMB lensing \citep{Hu2000}.  We advise readers to consult Ref.~\cite{Lewis_Challinor2006} for a complete and rigorous review of recent advancements on the theory of CMB lensing and \cite{Bartelmann_Schneider2001} for a general review of gravitational weak lensing.

\subsection{CMB Lensing---Flat-sky approximation}
\label{ssec:lensing}

We consider a lensed CMB temperature anisotropy in the direction $\hat{n}$ on the sky, $\tilde{\Theta}(\hat{n})$, and an unlensed temperature anisotropy $\Theta(\hat{n} + \alpha)$ where  $\alpha$ is the deflection angle due to a source with lensing potential $\psi(\hat{n})$, $\alpha \equiv \nabla\psi(\hat{n})$. $\tilde{\Theta}(\hat{n})$ can be expanded as
\begin{eqnarray}
	\label{eq:lensedT}
	\nonumber \tilde{\Theta}(\hat{n}) &=& \Theta(\hat{n}) + \nabla_{i}\psi \nabla^{i} \Theta(\hat{n}) \\ 
	&& + \frac{1}{2} \nabla_{i}\psi \nabla_{j}\psi \nabla^{i} \nabla^{j} \Theta(\hat{n}) + \ \mathcal{O}(\psi^3).
\end{eqnarray}
The Fourier transform of Eq.~(\ref{eq:lensedT}) is
\begin{equation}
	\label{eq:FourierT}
	\tilde{\Theta}(\Bell) = \Theta(\Bell) - \int\frac{\diff^2 \ell_1}{(2\pi)^2}\Theta(\Bell_1) L(\Bell, \Bell_1),
\end{equation}
where the lensing kernel $L(\Bell, \Bell_1)$ is given by
\begin{eqnarray}
	\label{eq:lensingkernel}
	\nonumber L(\Bell, \Bell_1) & = & \psi(\Bell - \Bell_1) (\Bell - \Bell_1) \cdot \Bell_1 \\
	\nonumber && - \frac{1}{2} \int \frac{\diff^2 \ell_2}{(2\pi)^2}\ \psi(\Bell_2) \psi(\Bell - \Bell_1 - \Bell_2) ( \Bell_1 \cdot \Bell_2) \\
	&& \times \left( \Bell_1 \cdot (\Bell - \Bell_1 - \Bell_2) \right).
\end{eqnarray}
$\Theta(\hat{n})$ is assumed Gaussianly distributed.  Therefore, the only independent correlation function is the two-point correlation function,
\begin{equation}
	\label{eq:2corr}
	\langle \Theta(\Bell)^* \Theta(\Bell^\prime)\rangle = (2\pi)^{2} \delta_D^{2}(\Bell - \Bell^\prime) C_{\ell}^{\Theta\Theta},
\end{equation}
where $\delta_D^{2}(\Bell - \Bell^\prime)$ is the 2D Dirac delta function and $C_\ell^{\Theta\Theta}$ is the $\Theta\Theta$-multipole moment of the order $\ell$.  From Eqs.~(\ref{eq:FourierT})--(\ref{eq:2corr}),
\begin{eqnarray}
	\label{eq:cltem}
	\nonumber\tilde{C}_{\ell}^{\Theta\Theta} &=& C_{\ell}^{\Theta\Theta} \left[ 1 - \int \frac{\diff^2\ell_1}{(2\pi)^2} \left( \Bell \cdot \Bell_1 \right)^2 C_{\ell_1}^{\psi\psi}\right] \\
	\nonumber&&+ \int\frac{\diff^2 \ell_1}{(2\pi)^2}\ \left( \Bell_1 \cdot \left( \Bell - \Bell_1\right) \right)^2 \left[ C_{\ell_1}^{\Theta\Theta} C_{\vert \Bell - \Bell_1\vert}^{\psi\psi} + C_{\ell_1}^{\Theta\psi} C_{\vert\Bell - \Bell_1\vert}^{\Theta\psi} \right].\\
\end{eqnarray}
The first term in Eq.~(\ref{eq:cltem}) could be interpreted as a transfer of the angular power spectrum on scale $\Bell$ into lensing scale $\Bell_1$ while the second term is a consequence of the convolution of $\Theta$ power spectra with the lensing power spectra.  Our result is consistent with Ref.~\cite{Hu2000} except for an inclusion of the temperature anisotropy and lensing potential cross-correlation $C_{\ell}^{\Theta\psi}$.

\subsection{Covariance matrix and Fisher analysis}
\label{ssec:cov}

\begin{figure}\centering
	\includegraphics[scale=0.65]{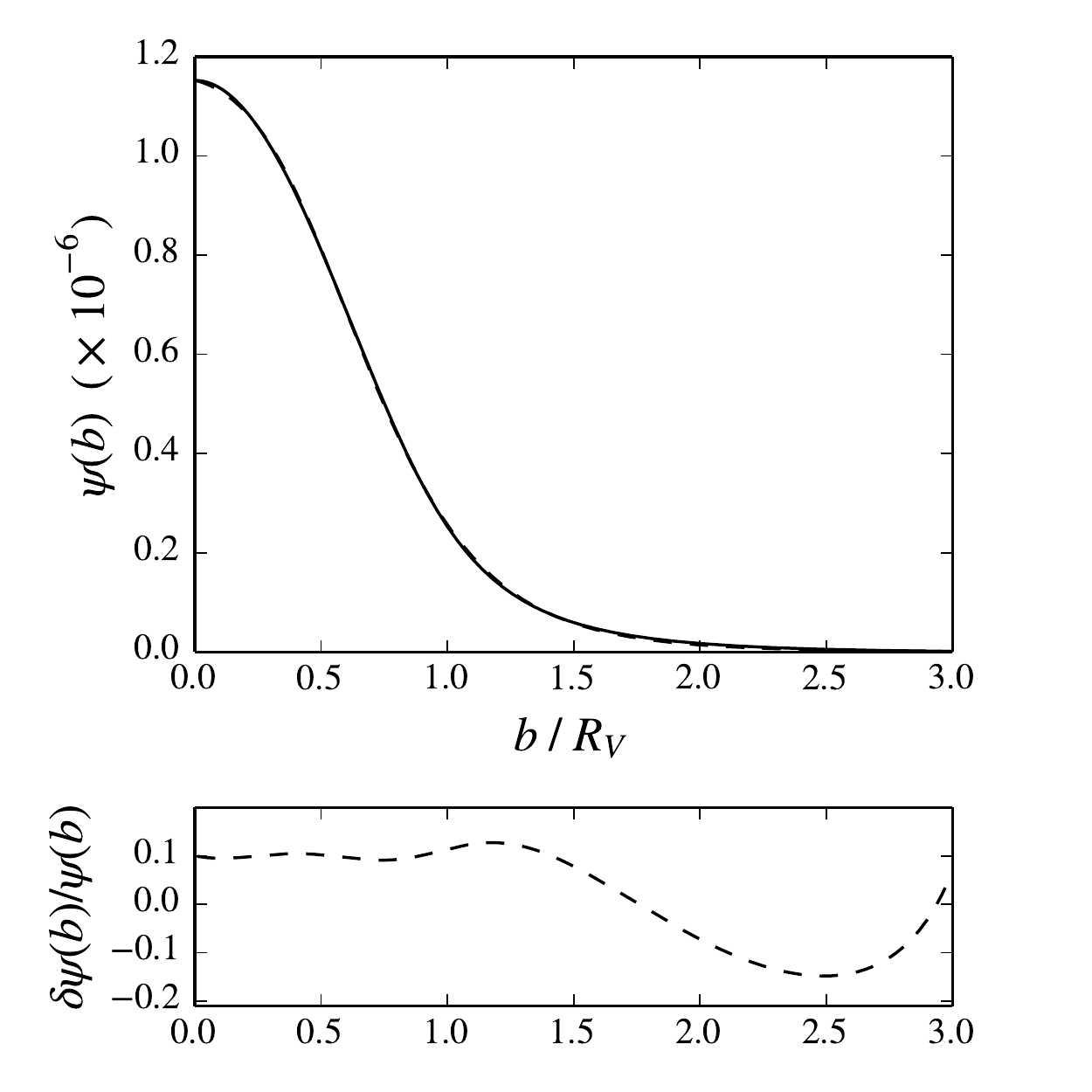}
	\caption{\label{fig2} (top panel) The void lensing potential for $R_V = 30.0$ Mpc $h^{-1}$ at $z = 0.5$ (solid) and the analytical fitting function Eq.~(\ref{eq:fitpotential}) (dashed).  (bottom panel) The fractional difference between the analytical fitting function and the lensing potential calculated numerically.}
\end{figure}

In order to forecast the ability of a given survey to constrain cosmological parameters, we adopt the Fisher matrix formalism \citep{Tegmark_ea1997}.  The CMB lensing covariance matrices formalism is adapted from Ref.~\cite{Benoit-Levy_ea2012} and the bandpower estimator from Ref.~\cite{Smith_ea2006}.  The bandpower estimator for lensed temperature anisotropies is given by
\begin{equation}
	\label{eq:estimator}
	\Delta^{\tilde{\Theta}\tilde{\Theta}}_i = \frac{1}{4\pi f_{\scriptsize{\mbox{sky}}} \alpha_i} \int_{\Bell\ \in\ i}\diff^2 \ell\ \left(\frac{\ell^2}{2\pi}\right) \tilde{\Theta}^*(\Bell) \tilde{\Theta}(\Bell),
\end{equation}
where $f_{\scriptsize{\mbox{sky}}}$ is the fraction of the sky covered by the survey.
\begin{equation}
	\alpha_i = \int_{\Bell\ \in\ i} \diff^2\ell,
\end{equation}
is the integrated $\ell$-space area of the $i$th band power.  In this article, we only consider the temperature anisotropy.  From the estimator in Eq.~(\ref{eq:estimator}), the covariance matrix for temperature anisotropy autocorrelation is 
\begin{eqnarray}
	\label{eq:covmatrix}
	\mbox{Cov}(\Delta^{\tilde{\Theta}\tilde{\Theta}}, \Delta^{\tilde{\Theta}\tilde{\Theta}})_{ij} &=& \langle \Delta^{\tilde{\Theta}\tilde{\Theta}}_i \Delta^{\tilde{\Theta}\tilde{\Theta}}_j \rangle - \langle \Delta^{\tilde{\Theta}\tilde{\Theta}}_i \rangle \langle \Delta^{\tilde{\Theta}\tilde{\Theta}}_j \rangle,
\end{eqnarray}
The indices $i, j$ refer to bins in $\ell$-space.  The full expression for $\mbox{Cov}(\Delta^{\tilde{\Theta}\tilde{\Theta}}, \Delta^{\tilde{\Theta}\tilde{\Theta}})_{ij}$ is given in Appendix~\ref{sec:appendixA}.  We assume no cross-correlation between $\Theta$ and $\psi$ for voids.  In term of the covariance matrix, the Fisher matrix is given by
\begin{equation}
	\label{eq:fisher}
	F_{\alpha\beta} = \left(\frac{\partial}{\partial p_\alpha}\langle\Delta^{\tilde{\Theta}\tilde{\Theta}}\rangle\right)^T\left(\mbox{Cov}(\Delta^{\tilde{\Theta}\tilde{\Theta}}, \Delta^{\tilde{\Theta}\tilde{\Theta}})\right)^{-1}\left(\frac{\partial}{\partial p_\beta}\langle\Delta^{\tilde{\Theta}\tilde{\Theta}}\rangle\right),
\end{equation}
where $p_\alpha$ and $p_\beta$ are cosmological parameters on which the bandpower depends. $\displaystyle \partial\ \langle\Delta^{\tilde{\Theta}\tilde{\Theta}}\rangle/\partial p_\alpha$ is a column vector of the partial derivative of $\langle\Delta^{\tilde{\Theta}\tilde{\Theta}}\rangle$ with respect to the parameter $p_\alpha$ as explained in details in Ref.~\cite{Chantavat_ea2011}.

\section{Methods}
\label{sec:methods}

We now forecast the sensitivity of CMB lensing of voids on the temperature angular power spectrum of the CMB $C_\ell$ on the surveys.

\subsection{Void model}
\label{ssec:voidmodel}

For most voids, the underdense central region is surrounded by an external overdense region called a \textit{compensation}.  The recent simulations of Ref.~\cite{Hamaus_ea2014b} have shown that the radial profile of \textit{averaged} voids is spherically symmetric and is well fitted empirically by
\begin{equation}
	\label{eq:voidprofile}
	\rho_V(r)/\bar{\rho}_M =  1 + \delta_c \frac{1 - (r / R_S)^{\alpha}}{1 + (r / R_V)^{\beta}},
\end{equation}
where $\bar{\rho}_M$ is the mean cosmic matter density and $R_V$ are the characteristic void radius.  $R_S$ is a scale radius where $\rho_V = \bar{\rho}_M$.  We shall take the parameters as $R_S/R_V = 0.93$, $\alpha = 2.13$, $\beta = 9.24$ and $\delta_c = -0.85$ for $R_V$ within 20 -- 60 Mpc $h^{-1}$ \citep{Hamaus_ea2014b}.  The choice of parameters is made such that the voids are well compensated.  Even though voids, in general, do not have a spherical shape as in the stacked void profile, we shall take the average over many voids with different ellipticities and orientations as our approximation \citep{Pisani_ea2014}.

\begin{figure}\centering
	\includegraphics[scale=0.5]{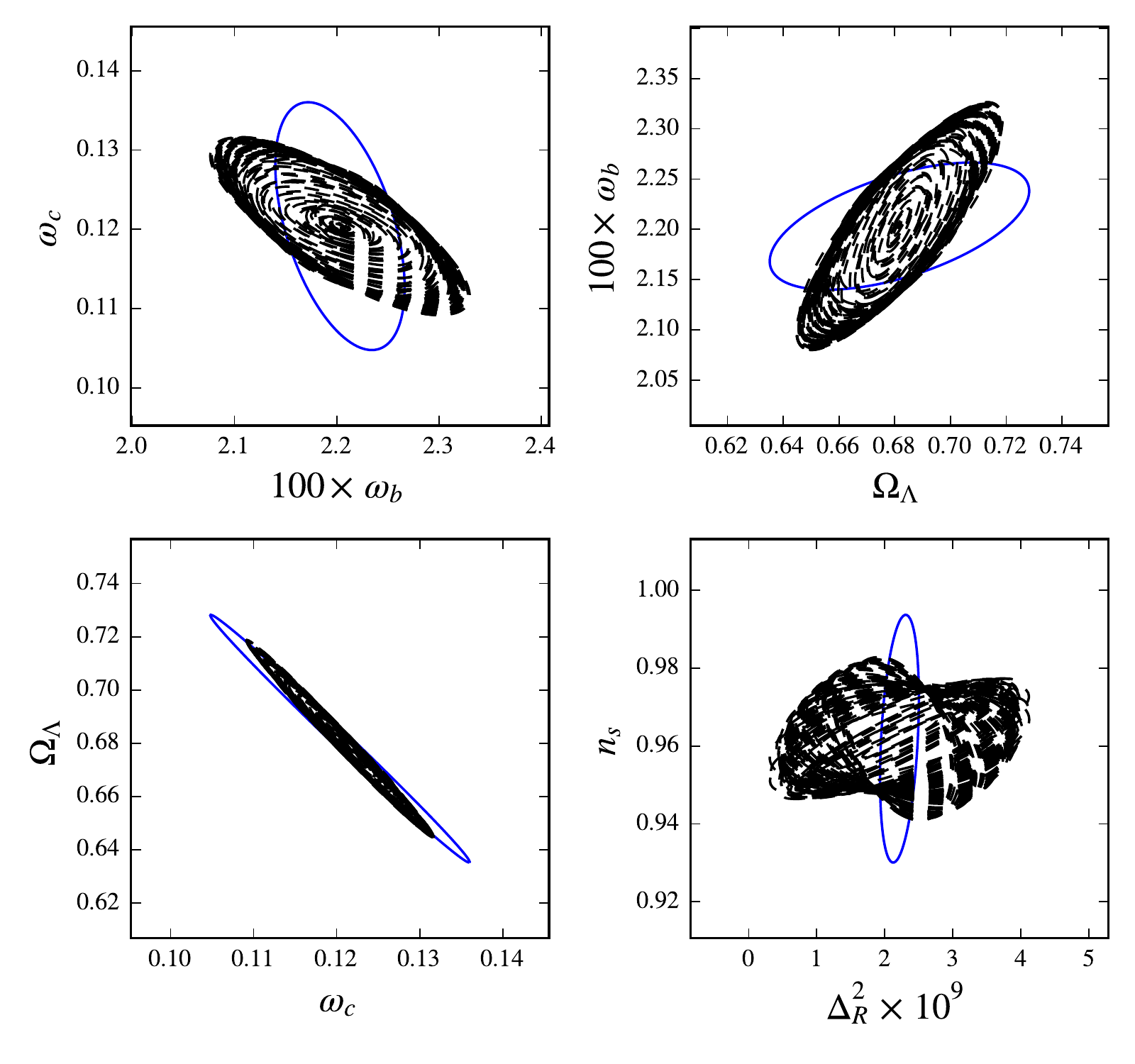}
	\caption{\label{fig3} 95\% confidence level constraints on some of the cosmological parameter pairs: $100\ \omega_b$ and $\omega_c$ (top-left),  $\Omega_\Lambda$ and $100\  \omega_b$ (top-right), $\omega_c$ and $\Omega_\Lambda$ (bottom-left) and $\Delta_{\mathcal{R}}^2$ and $n_S$ (bottom-right) for $N_V = 5$ with random sky (solid) and multiple realizations of void populations (dashed).   The scatter on constraints with different realizations is due to the sensitivity of the lensing potential with $R_V$ and $N_V$.}
\end{figure}

For a weak gravitational field and a perfect fluid assumption, the distortion of spacetime is caused by the Newtonian gravitational potential $\Psi_{\scsc{N}}$ which obeys the Poisson equation,
\begin{equation}
	\label{eq:poisson}
	\nabla^2 \Psi_{\scsc{N}} = 4 \pi G \bar{\rho}_M (1 + z) D_{+}(z)\delta_M(z = 0),
\end{equation}
where $\boldsymbol{\nabla}$ is the comoving gradient operator.  $D_{+}(z)$ is the linear growth function normalized to unity at the present epoch, and $z$ is the redshift.  The gravitational lensing potential $\psi(\hat{n})$ is given by
\begin{eqnarray}
	\psi(\hat{n}) = -\frac{2}{c^2} \int \mbox{d}\chi\ \boldsymbol{\nabla}_\perp \Psi_{\scsc{N}}(\chi\hat{n}),
\end{eqnarray}
where $\chi$ is the comoving distance to the lensing source.  $\nabla_\perp$ is the transverse derivative.  The integral is performed along the line of sight.  Similarly, in term of angular separation $\boldsymbol{\theta}$,
\begin{eqnarray}
	\label{eq:totalpsi}
	\psi(\boldsymbol{\theta}) &=& \int\diff^2 \hat{n}\ \left[ \sum_{i}^{N_V} \delta_D^{2}\left( \hat{n} - \hat{n}_i \right) \psi_i(\hat{n}_i; R_{V, \; i}, z_i)\right],
\end{eqnarray}
where $N_V$ is the number of voids.  $\hat{n}_i$'s are the positions of voids in the sky. The Fourier transform of the lensing potential into $\ell$-space is given by
\begin{equation}
	\label{eq:psil}
	\psi(\Bell; R_V, z) = \int \diff^2 \theta\ \psi(\boldsymbol{\theta}; R_V, z) \exp\left( -i \Bell \cdot \boldsymbol{\theta} \right).
\end{equation}
We would advise the reader \cite{Amendola_ea1999} on detailed calculation of the lensing potential from the Newtonian gravitational potential. Figure.~\ref{fig1} shows the lensing potentials of voids and their corresponding angular power spectra. The lensing potential in real space with voids as a function of the impact parameter $\boldsymbol{b} \equiv D_K \boldsymbol{\theta}$, where $D_K$ is the comoving angular diameter distance, is well approximated by the function
\begin{equation}
	\label{eq:fitpotential}
	\psi(b; R_V, z) = \mathcal{S}(R_V, z)\times\ \tilde{\psi}(b / R_V),
\end{equation}
where $\tilde{\psi}(x)$ is the scale-invariant lensing potential and $\mathcal{S}(R_V, z)$ is the lensing potential scaling factor.
\begin{equation}
	\tilde{\psi}(x) = \psi_0 \exp\left(\Gamma_0 x^{\gamma_0}\right) \times\ \left(1.0 +  x^{\gamma_1}\right)^{\gamma_2},
\end{equation}
where $\psi_0 = 9.06\times 10^{-2}$ Mpc$^{2}$ h$^{-2}$, $\gamma_0 = 1.29$, $\gamma_1 = 2.86$, $\gamma_2 = -1.72$, and $\Gamma_0 = -0.31$.
\begin{eqnarray}
	\nonumber\mathcal{S}(R_V, z) &=& \frac{16\pi G}{c^2} \Omega_M \bar{\rho}_{c} \left(\frac{R_V}{\mbox{Mpc}\ h^{-1}}\right)^3 \times \frac{(1 + z)^3 D_{+}(z)}{(D_K(z) / \mbox{Mpc}\ h^{-1})}, \\
\end{eqnarray}
where $\bar{\rho}_c$ is the critical density at the present epoch. Our fitting function for the lensing potential is accurate within $\sim$10\% over the range well within $3 R_V$ (see Fig.~\ref{fig2}).

\begin{figure}\centering
	\includegraphics[scale=0.5]{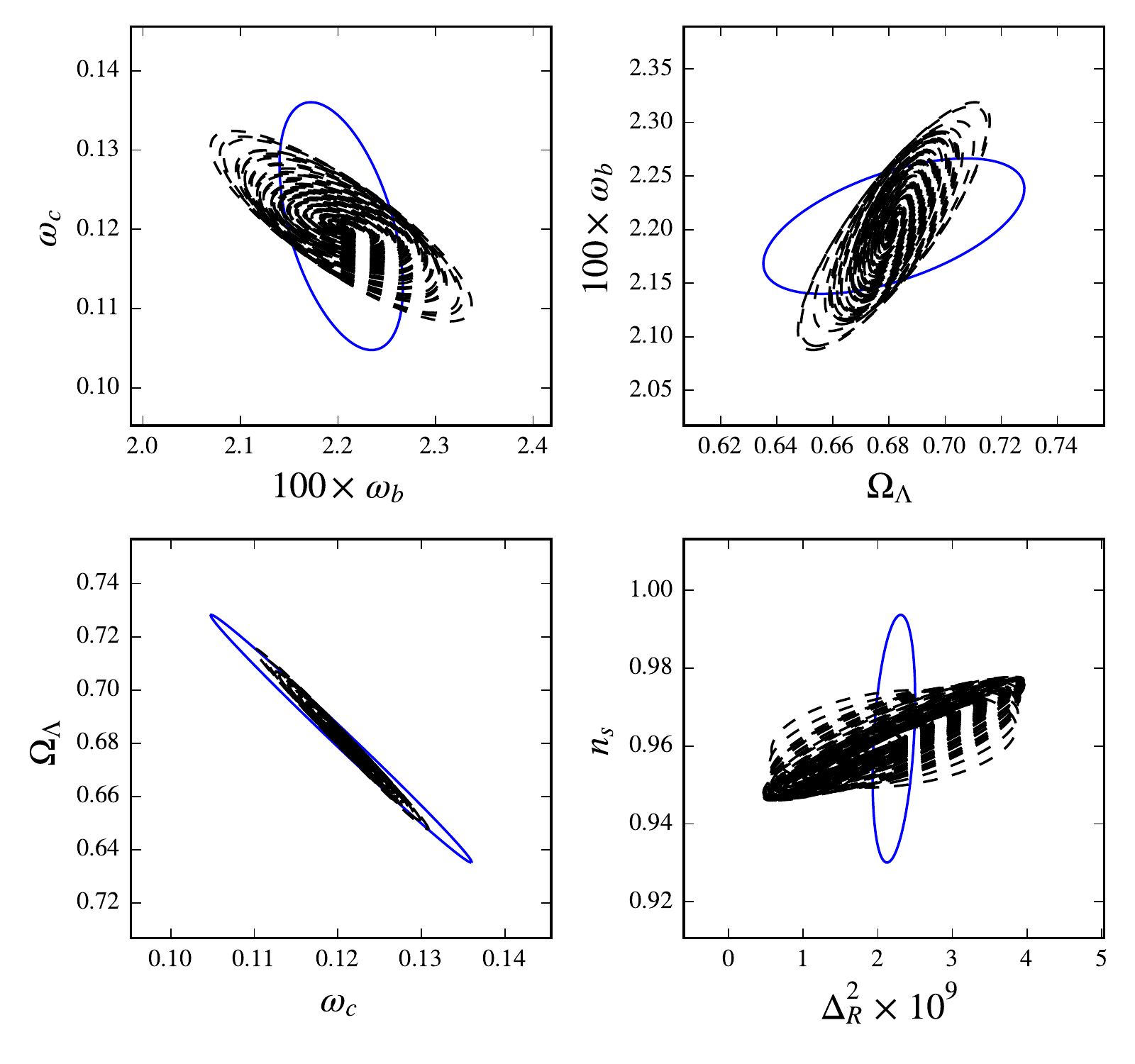}
	\caption{\label{fig4} Same as Fig.~\ref{fig3} but with $N_V = 10$.}
\end{figure}

\subsection{Void distribution}

In order to give an estimate of the void distribution as a function of the radius along the line of sight, the number density of voids is needed \citep{Jennings_ea2013, Sheth_vandeWeygaert2004, Sutter_ea2014c}.  However, for our forecast on the CMB lensing signal with voids, we assume the void number function for a EUCLID-like mission based on \cite{Sheth_vandeWeygaert2004}
\begin{eqnarray}
	\label{eq:nvoid}
	n_V(M) &=& \frac{\bar{\rho}_M}{M^2} \nu f(\nu) \frac{\mbox{d}\ln \nu}{\mbox{d}\ln M},
\end{eqnarray}
where $M$ is the void mass and $\nu = \delta_v^2 / \sigma^2(M)$ with $\delta_v$ being the critical underdensity for the void and $\sigma^2(M)$ is the variance of the density field.
\begin{equation}
	\nu f(\nu) = \sqrt{\frac{\nu}{2\pi}} \exp\left( -\frac{\nu}{2}\right) \exp\left( -\frac{|\delta_c|}{\delta_v}\frac{\mathcal{D}^2}{4\nu}-2\frac{\mathcal{D}^4}{\nu^2}\right),
\end{equation}
where $\mathcal{D} \equiv |\delta_v| / (\delta_c + |\delta_v|)$ and $\delta_c = 1.686$.  We take $\delta_v = -0.43$ from the HOD dense simulation in Ref.~\cite{Sutter_ea2014d}.  The radius distribution of voids in one-dimensional space will be $\sim n_V(R_V) D_K(z)^2 \times 1.0^{\circ} \times 1.0^{\circ}$ for a squared degree patch where $R_V = 1.7\times(3 M / 4 \pi \bar{\rho}_M)^{1/3}$.  At this stage we are not considering several practical difficulties which may complicate the recognition of voids in the surveys and assume that the surveys can identify voids down to the characteristic size of $R_V \sim$ 20 Mpc $h^{-1}$ for our fiducial surveys within the redshift range.  We select voids of $R_V > 20$ Mpc $h^{-1}$ as indicated in Ref.~\cite{Hamaus_ea2014a}, a transition radius from overcompensated to undercompensated voids.  The undercompensated voids tend to inhibit in the underdense region of the Universe where our lines of sight are chosen.  The determination of the void radius is subjected to the uncertainty in mapping the galaxies to the underlying dark matter \citep{Sutter_ea2014a}.  In this analysis, we assume 10\% statistical uncertainty in $R_V$ measurement which will be marginalized over the cosmological parameters.

We shall model how the centers of the voids are misaligned along the line of sight by allowing centers of voids to be offset uniformly within a field of view in Eq.~(\ref{eq:totalpsi}).  As small voids are commonly found in overdensed structures, larger voids are more abundant when we select patches of the sky which are free of clusters from low-$z$ cluster surveys.  Given a preselected patch of the sky with no clusters found in low-$z$ surveys, the chance of encountering sizeable clusters to the field of view at higher redshift is assumed negligible.  The distribution of voids is assumed Poissonian; therefore the lensing effect of voids whose centre are out of the field of view are averaged out.  In addition, we assume a nominal $f_{\mbox{\scriptsize{patch}}}$ of $1.0^{\circ} \times 1.0^{\circ}$ such that voids with $R_V > 20$ Mpc $h^{-1}$ could be well observed within the patch from $z = 0.0 - 1.0$.

We can express the lensing potential of voids as
\begin{equation}
	\label{eq:voidtotal}
	\psi_{\mbox{\scriptsize total}}(\boldsymbol{\theta}) = \sum_{j}^{N_V}\psi_{j}(\boldsymbol{\theta} - \boldsymbol{\theta}_j),
\end{equation} 
where $\psi_j(\boldsymbol{\theta})$ is the lensing potential of $j$th void and $\boldsymbol{\theta}_{j}$ is the center of the $j$th void from the common center.  The contribution to the angular power spectrum due to the lensing effect of voids is given by
\begin{equation}
	\label{eq:clmanyvoids}
	C_{\ell, \mbox{\scriptsize total}}^{\psi\psi} = \sum_{j}^{N_V} C^{\psi\psi}_{\ell, j} +  2 \sum_{j < k}^{N_V} J_0(\ell \Delta\theta_{jk}) \left\langle \psi_{j}(\boldsymbol{\ell})\psi^{*}_{k}(\boldsymbol{\ell})\right\rangle,
\end{equation}
where $\Delta\boldsymbol{\theta}_{jk} \equiv \boldsymbol{\theta}_{j} - \boldsymbol{\theta}_{k}$ and $J_n(x)$ is the Bessel function of the first kind. The first term is the correlation from the same void and the second term is the correlation due to different voids. The detail derivation for Eq.~(\ref{eq:clmanyvoids}) is given in Appendix~\ref{sec:appendixB}. \\

To summarize our method, we shall proceed as follows:
\begin{itemize}
	\item Generate 100 realizations of a sky patch of $1.0^{\circ} \times 1.0^{\circ} $ square degree with voids distributed along a line of sight given in terms of $R_V$ and $z$ for $N_V = 5, 10$ taking the misalignment into account.
	\item The lensing potential in Eq.~(\ref{eq:psil}) is calculated from the void profile [Eq.~(\ref{eq:voidprofile})] for each void in a given realization.  The resulting void lensing potentials in a line-of-sight are combined in Eq.~(\ref{eq:clmanyvoids}) for $C_{\ell, \mbox{\scriptsize total}}^{\psi\psi}$ in the line of sight.
	\item Calculate the covariance matrices [Eq.~(\ref{eq:covmatrix})] and the Fisher matrices [Eq.~(\ref{eq:fisher})], and get the parameter constraints with the void parameters, ($\alpha$, $\beta$, $\delta_c$, $r_s/r_v$) and $R_V$ as nuisance parameters to be marginaliszed with a 10\% prior on $R_V$.
\end{itemize}

\section{Results}
\label{sec:results}

In this article, we shall assume a noise-free small-area CMB observation on a preselected part of the sky where multiple voids are found by large-scale structure surveys such as BigBOSS \citep{BigBOSS2009}, DES \citep{DES}, LSST \citep{LSST2009} and EUCLID \citep{Laureijs_ea2011}.  We also assume the accurate determination of the dark matter void radius to $\sim10$\% level, which will be included in the Fisher analysis.  In addition, we assume a void profile by Ref.~\cite{Hamaus_ea2014b} where void parameters are chosen such that voids are well compensated.  Even though most voids are not compensated, they are inclined to be \textit{undercompensated} for voids with $R_V > 20\ \mbox{Mpc}\ h^{-1}$ \citep{Hamaus_ea2014b}.  Hence, we include void parameters in the analysis as nuisance parameters.

As an illustrative demonstration of the importance of the gravitational lensing by voids on cosmological parameter constraints, we shall take $\omega_b$~and~$\omega_c$, $\omega_b$~and~$\Omega_\Lambda$, $\omega_c$~and~$\Omega_\Lambda$ and $\Delta_R^2$~and~$n_s$ pairs as an example shown in Fig.~\ref{fig3}~and~\ref{fig4}.  In both figures, 100 realizations of voids with radius 20--60 Mpc $h^{-1}$ within redshift 0.0--1.0 are generated according to the void number functions by Ref.~\cite{Sheth_vandeWeygaert2004}.  The constraints vary significantly due to the random nature of the distributions.  However, the degeneracy directions are significantly different from an arbitrary sky patch.

\begin{figure}\centering
	\includegraphics[scale=0.5]{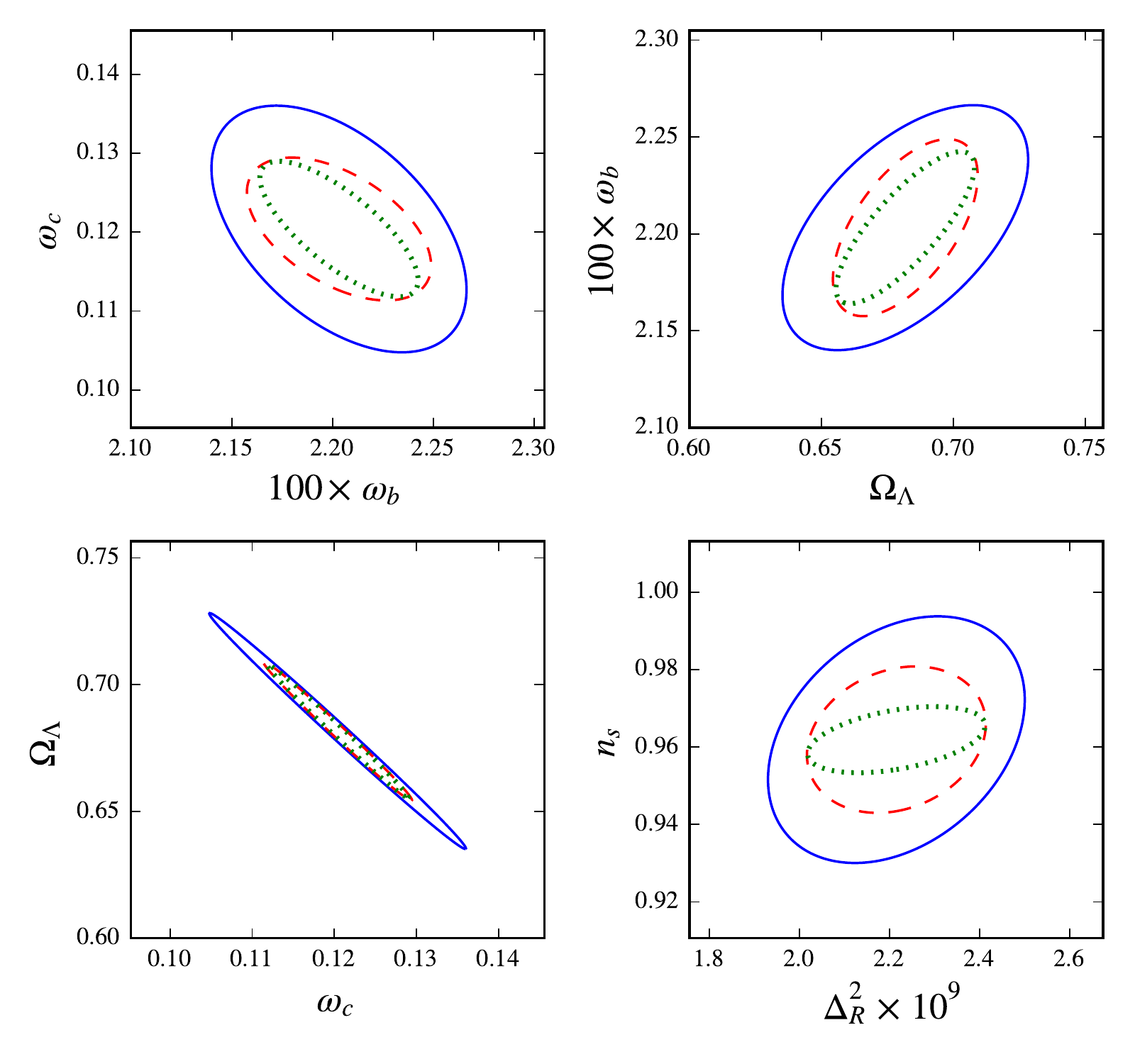}
	\caption{\label{fig5} 95\% confidence level constraints on some of the cosmological parameter pairs; $100\ \omega_b$ and $\omega_c$ (top-left), $\Omega_\Lambda$ and $100\  \omega_b$ (top-right), $\omega_c$ and $\Omega_\Lambda$ (bottom-left) and $\Delta_{\mathcal{R}}^2$ and $n_S$ (bottom-right) for a square-degree random sky (solid), random sky + $N_V = 5$ (dashed) and random sky + $N_V = 10$ (dotted).}
\end{figure}

The full parameters constraint are shown in Table.~\ref{table:params} where we choose the median of the ellipses as a representation of the realizations for $N_V = 5$ and $10$ in Fig.~\ref{fig5}.  The constraints on void parameters are given where applicable.  We also provide combined parameter constraints between an arbitrary square-degree sky patch and a square-degree sky patch with voids.

{\renewcommand{\arraystretch}{1.5}
\begin{table*}\centering
	\caption{\label{table:params} 68\% C.L. parameter constraints on the cosmological parameters.}
	\begin{ruledtabular}
	\begin{tabular}{l|c|c|c|c|c|c|c|c|c|c}
		& $100 \times \sigma_{\omega_b}$ &$\sigma_{\omega_c}$ &$\sigma_{\Omega_\Lambda}$ & $\sigma_{\Delta_{\mathcal{R}}^2} \times 10^9$ & $\sigma_{n_s}$ & $\sigma_{\tau}$ & $\sigma_{\alpha} \times 10^5$ & $\sigma_{\beta} \times 10^5$ & $\sigma_{\delta_c} \times 10^4$ & $\sigma_{r_s / r_v} \times 10^4$\\
		\hline
		Random & 0.0509 & 0.01258 & 0.03747 & 0.2296 & 0.02563  & 0.0597 & N/A & N/A & N/A & N/A \\
		\hline
		$N_{V} = 5$ & 0.2721  & 0.02283 & 0.07795 & 7.786  & 0.09667 & 1.747 & 2.176 & 7.741 & 4.886 & 1.990 \\
		$N_{V} = 10$ & 0.1139  & 0.00981 & 0.02939 & 6.604 & 0.05429 & 1.487 & 0.767 & 3.435 & 1.398 & 0.796 \\
		\hline
		Random + $N_{V} = 5$ & 0.0368 & 0.00729 & 0.02208 & 0.1599 & 0.01522 & 0.0395 & N/A & N/A & N/A & N/A \\
		Random + $N_{V} = 10$ & 0.0316 & 0.00691 & 0.02112 & 0.1588 & 0.00687 & 0.0388 & N/A & N/A & N/A & N/A \\
	\end{tabular}
	\end{ruledtabular}
\end{table*}
}
\section{Discussions and Conclusions}
\label{sec:diss}

The main advantage of CMB lensing by voids arises from the fact that $C_\ell^{\psi\psi}$ for voids scales approximately as $\sim N_{V}^2$ along the line of sight.  The scaling relation of void lensing power spectra comes from the linearity of the void lensing potential [see Eq.~(\ref{eq:voidtotal})].  Hence, the void power spectra are enhanced over the intrinsic CMB power spectra by $\sim N_V^2$.  However, the constraints are limited by the scatter in the void profile.   Another advantage is the sensitivity of $\displaystyle C_\ell^{\psi\psi}$ with $\displaystyle R_V$ (See Fig.~\ref{fig1}).  This implies that better constraints could be achieved with larger voids located at low redshift.  However, the chance of spoiling the lensing effect by Sunyaev-Zel’dovich (SZ) effects of intervening clusters of galaxies is possible.  The impact from SZ contamination is expected to be more important than the lensing caused by clusters: the typical angular extension, $\theta_{500}$, of the SZ temperature profile is a few $10'$ to $100'$ (see e.g.~Refs.~\cite{PlanckXXIX, Whitbourn_ea2014}).  Hence, the purity of the selected sky is important.

The assumption of finding a sizeable cluster at higher redshift is crucial in the analysis.  We use Ref.~\cite{Jenkins_ea2001}'s mass function and Ref.~\cite{Cooray_Sheth2002} to calculate a cluster of size $> 20$ Mpc h$^{-1}$ and find that the probability is $\lesssim 10^{-5}$, which is negligible.  In addition, some parameters have degeneracies lifted by incorporating the additional void information.  Furthermore, we assume that, regarding the angular size of the patch at a given redshift, the lensing effect of intervening galaxies id negligible.  The validity of our results relies on the search for such $1.0^{\circ} \times 1.0^{\circ}$ patches of the sky.

The assumed number function gives the mean radius of $\bar{R}_V \approx 23.2$ Mpc $h^{-1}$ in a low density part of the Universe.  The probability of finding the patch of the sky with 5 -- 10 voids is approximately $\sim10^{-5}$, which is equivalent to $\sim 1$ patch per universe.  However, our analysis only based on voids resides within redshift $0.0 - 1.0$, and hence the chance of finding such a patch would be greater for higher redshift.  We shall take our evaluation as a conservative estimate for finding such a patch. 

The constraints on cosmological parameters get improved where larger voids and smaller redshifts are added.  Not only does the area of the ellipse shrink, but also the degeneracy direction changes.  The change in the degeneracy direction reflects the fact that the intrinsic degeneracy direction of the voids power spectrum is different from the intrinsic CMB power spectrum.  This is clearly seen in the $n_s$ vs $\Delta_{R}^2$ constraint.  Even though our void profile does not have an explicit dependence on $\omega_b$, the improvement on $\omega_b$ is due to the fact that the lensed power spectra with voids are  convolution functions of the intrinsic CMB power spectra that depend on $\omega_b$.

The other secondary effect besides lensing is notably the SZ effect \citep{Zeldovich1968} and the Rees-Sciama (RS) effect \citep{Rees_Sciama1968}.  The SZ effect is expected not to have a sizeable contribution in an underdense region \citep{Birkinshaw1999}.  One would expect that there should be no SZ effect from voids at all as there should be no significant amount of gas.  The RS effect, however, may have a significant effect for very large voids, $\left | \delta T^{\mbox{\scriptsize RS}}/T \right | \propto \displaystyle R_V^{\beta}$ where $\beta \simeq 2.5-3.0$.  For a single void with $\displaystyle R_{V{\mbox{\scriptsize , eff}}}=\bar{R}_V \approx 23.2$ Mpc $h^{-1}$, the predicted $\ell(\ell+1)C_{\ell} ^{\psi\psi}/2\pi \approx 0.1 ~\mu \mbox{K}^2$ at $\ell \approx$ 100--200.  For a one square-degree patch with 10 of those voids in the slight line, lensing contribution becomes $\ell (\ell + 1) C_{\ell}^{\Theta\Theta} / 2\pi \approx 600 \mu K^2$.  A full-sky ray-tracing analysis by Ref.~\cite{Cai_ea2010} estimated the RS contribution to the CMB anisotropy $\ell(\ell+1)C_{\ell} ^{\mbox{\scriptsize RS}}/2\pi \approx 0.1 ~\mu \mbox{K}^2$ at the similar multipoles for redshift slice $0.17 < z < 0.57$ for both voids and clusters.  In this work, we therefore neglect the RS effect for the aforementioned reasons. A full ray-tracing analysis of weak lensing and other secondary anisotropies from voids will be the subject of our future investigation.

\begin{acknowledgements}
We would like to thank Sirichai Chongchitnan and Nico Hamaus for useful comments and Khamphee Karwan for his generous provision of computing facilities in numerically intensive parts of our calculation.  T.~C. acknowledges the support from the National Astronomical Research Institute of Thailand (NARIT) and Naresuan University Grant No. R2555C018.  This work is supported by a NARIT research grant and its High Performance Computer facility.  B.~W. acknowledges funding from an ANR Chaire d’Excellence (Grant No. ANR-10-CEXC-004-01) and the UPMC Chaire Internationale in Theoretical Cosmology.  This work has been done within the Labex Institut Lagrange de Paris (Reference No. ANR-10-LABX-63), part of the Idex SUPER, and received financial state aid managed by the Agence Nationale de la Recherche, as part of the Programme Investissements d’Avenir under Reference No. ANR-11-IDEX- 0004-02.
\end{acknowledgements}

\bibliography{voidLensing}

\appendix
\section{Covariance matrix for CMB lensing}
\label{sec:appendixA}

Following Ref.~\cite{Smith_ea2006}, we obtain the expression for the covariance matrix for CMB lensing,
\begin{eqnarray}
	\label{eq:covmatrix}
	\nonumber\mbox{Cov}(\Delta^{\tilde{\Theta}\tilde{\Theta}}, \Delta^{\tilde{\Theta}\tilde{\Theta}})_{ij} &=& \langle \Delta^{\tilde{\Theta}\tilde{\Theta}}_i \Delta^{\tilde{\Theta}\tilde{\Theta}}_j \rangle - \langle \Delta^{\tilde{\Theta}\tilde{\Theta}}_i \rangle \langle \Delta^{\tilde{\Theta}\tilde{\Theta}}_j \rangle, \\
	&=& \mathcal{G}_{i} \delta_{ij} + \mathcal{H}_{i} \delta_{ij} + \mathcal{I}_{ij} + \mathcal{J}_{ij},
\end{eqnarray}
where the indices $i, j$ refer to bins in $\ell$-space and $\delta_{ij}$ is the Kronecker's delta. $\mathcal{G}_{i}$ is the Gaussian term and the other terms are the non-Gaussian parts of the covariance matrix.  The Gaussian term is given by
\begin{equation}
	\mathcal{G}_{i} = \frac{2(2\pi)^2}{4\pi f_{\scriptsize{\mbox{sky}}} \alpha_i^2} \int_{\Bell \in i} \mbox{d}^2 \ell\ \left(\frac{\ell^2}{2\pi}\right)^2 C_{\ell}^{\Theta\Theta}C_{\ell}^{\Theta\Theta}.
\end{equation}
The other terms are given by
\begin{widetext}
\begin{eqnarray}
	\mathcal{H}_{i} &=& \frac{4}{4\pi f_{\scriptsize{\mbox{sky}}} \alpha_i^2} \int_{\Bell \in i} \diff^2 \ell\ \ell^4  \int \frac{\diff^2 \ell_1}{(2\pi)^2}\ \Bigg[ C_{\ell}^{\Theta\Theta} \left( C_{\ell_1}^{\Theta\Theta} C_{\vert\Bell - \Bell_1\vert}^{\psi\psi} + C_{\ell_1}^{\Theta\psi} C_{\vert\Bell - \Bell_1\vert}^{\Theta\psi}\right) \left(\left(\Bell - \Bell_1\right) \cdot\Bell_1\right)^2 -\ C_{\ell}^{\Theta\Theta} C_{\ell}^{\Theta\Theta} C_{\ell_1}^{\psi\psi} \left( \Bell \cdot \Bell_1 \right)^2 \Bigg], \\
	\nonumber&&\mbox{}\\
	\mathcal{I}_{ij} &=& -\frac{2}{4\pi f_{\scriptsize{\mbox{sky}}} \alpha_i \alpha_j} \int_{\Bell \in i} \diff^2 \ell\ \int_{\Bell^{\prime} \in j} \diff^2 \ell^{\prime}\ \left( \frac{\ell^2}{2\pi} \right) \left( \frac{\ell^{\prime 2}}{2\pi} \right) \times \ \left(C_{\ell}^{\Theta\Theta} C_{\ell^\prime}^{\Theta\psi} C_{\ell^\prime}^{\Theta\psi} + C_{\ell^\prime}^{\Theta\Theta} C_{\ell}^{\Theta\psi} C_{\ell}^{\Theta\psi} \right) \left( \Bell \cdot \Bell^{\prime} \right)^2,\\
	\nonumber\mathcal{J}_{ij} &=& \frac{1}{4\pi f_{\scriptsize{\mbox{sky}}} \alpha_i \alpha_j} \int_{\Bell \in i} \diff^2 \ell\ \int_{\Bell^{\prime} \in j} \diff^2 \ell^{\prime}\ \left( \frac{\ell^2}{2\pi} \right) \left( \frac{\ell^{\prime 2}}{2\pi} \right) \times\ \Bigg( 2 \alpha_{+}\left(\Bell, \Bell^{\prime}\right) \mathcal{M}_{+}\left(\Bell, \Bell^{\prime}\right) \mathcal{M}_{+}\left(\Bell^\prime, \Bell\right) \\
	\nonumber&& +\  2 \alpha_{-}\left(\Bell, \Bell^{\prime}\right) \mathcal{M}_{-}\left(\Bell, \Bell^{\prime}\right) \mathcal{M}_{-}\left(\Bell^\prime, \Bell\right) +\ \beta_{+}\left(\Bell, \Bell^{\prime}\right) \mathcal{M}_{+}\left(\Bell, \Bell^{\prime}\right)^2 + \beta_{+}\left(\Bell^{\prime}, \Bell \right) \mathcal{M}_{+}\left(\Bell^{\prime}, \Bell\right)^2 \\
	&&+\ \beta_{-}\left(\Bell, \Bell^{\prime}\right) \mathcal{M}_{-}\left(\Bell, \Bell^{\prime}\right)^2 + \beta_{-}\left(\Bell^{\prime}, \Bell \right) \mathcal{M}_{-}\left(\Bell^{\prime}, \Bell\right)^2 \Bigg)
\end{eqnarray}
where
\begin{eqnarray}
	\mathcal{M}_{\pm}\left( \Bell, \Bell^{\prime}\right) &=& \left( \Bell \pm \Bell^{\prime} \right) \cdot \Bell,\\
	\alpha_{\pm}\left( \Bell, \Bell^{\prime}\right) &=& C_{\ell}^{\Theta\Theta} C_{\ell^\prime}^{\Theta\Theta} C_{\vert\Bell\pm\Bell^{\prime}\vert}^{\psi\psi} + C_{\ell}^{\Theta\Theta} C_{\ell^\prime}^{\Theta\psi} C_{\vert\Bell\pm\Bell^{\prime}\vert}^{\Theta\psi}\ C_{\ell^\prime}^{\Theta\Theta} C_{\ell}^{\Theta\psi} C_{\vert\Bell\pm\Bell^{\prime}\vert}^{\Theta\psi} + C_{\ell^\prime}^{\Theta\psi} C_{\ell}^{\Theta\psi} C_{\vert\Bell\pm\Bell^{\prime}\vert}^{\Theta\Theta},\\
	\beta_{\pm}\left( \Bell, \Bell^{\prime}\right) &=& C_{\ell}^{\Theta\Theta} C_{\ell}^{\Theta\Theta} C_{\vert\Bell\pm\Bell^{\prime}\vert}^{\psi\psi} + C_{\ell}^{\Theta\Theta} C_{\ell}^{\Theta\psi} C_{\vert\Bell\pm\Bell^{\prime}\vert}^{\Theta\psi}\ C_{\ell}^{\Theta\Theta} C_{\ell}^{\Theta\psi} C_{\vert\Bell\pm\Bell^{\prime}\vert}^{\Theta\psi} + C_{\ell}^{\Theta\psi} C_{\ell}^{\Theta\psi} C_{\vert\Bell\pm\Bell^{\prime}\vert}^{\Theta\Theta}.
\end{eqnarray}
\end{widetext}

Our result is consistent with Ref.~\cite{Smith_ea2006} except for the inclusion of the temperature anisotropy and lensing potential cross-correlation function.  In addition, we find correction terms due to the second order expansion in Eq.~(\ref{eq:lensingkernel}).

\section{Angular power spectrum for multiple nearly aligned lensing sources}
\label{sec:appendixB}

Suppose that we have $N$ number of lensing sources slightly misaligned along a line of sight; we can write the total lensing potential as
\begin{equation}
	\psi_{\mbox{\scriptsize total}}(\boldsymbol{\theta}) = \psi_{1}(\boldsymbol{\theta} - \boldsymbol{\theta}_1) + \ldots + \psi_{N}(\boldsymbol{\theta} - \boldsymbol{\theta}_N),
\end{equation}
where $\psi_j(\boldsymbol{\theta})$ is the lensing potential of $j$th lensing source and $\boldsymbol{\theta}_{j}$ is the center of the $j$th source from the common center.  The Fourier transform of the whole system will be
\begin{eqnarray}
	\nonumber \psi_{\mbox{\scriptsize total}}(\boldsymbol{\ell}) &=& \int \mbox{d}^2\theta\ \psi_{\mbox{\scriptsize total}}(\boldsymbol{\theta}) \exp\left(-i \boldsymbol{\ell} \cdot \boldsymbol{\theta}\right), \\
	\nonumber & = & \int \mbox{d}^2\theta\ \sum_{j} \psi_{j}(\boldsymbol{\theta} - \boldsymbol{\theta}_j) \exp\left(-i \boldsymbol{\ell} \cdot \boldsymbol{\theta}\right), \\
	\psi_{\mbox{\scriptsize total}}(\boldsymbol{\ell}) & = & \sum_{j} \int\mbox{d}^2\theta\ \psi_{j}(\boldsymbol{\theta}) \exp\left(-i \boldsymbol{\ell} \cdot (\boldsymbol{\theta} + \boldsymbol{\theta}_j)\right).
\end{eqnarray}

Since the angles ($\boldsymbol{\ell} \cdot \boldsymbol{\theta}$) and ($\boldsymbol{\ell} \cdot \boldsymbol{\theta}_{j}$) are independent (clearly shown if we express them in Cartesian coordinates), then
\begin{eqnarray}
	\nonumber\psi_{\mbox{\scriptsize total}}(\boldsymbol{\ell}) &=& \sum_{j} \exp\left(-i \boldsymbol{\ell} \cdot \boldsymbol{\theta}_j\right)\int\mbox{d}^2\theta\ \psi_{j}(\boldsymbol{\theta}) \exp\left(-i \boldsymbol{\ell} \cdot \boldsymbol{\theta}\right), \\
	&=& \sum_{j} \exp\left(-i \boldsymbol{\ell} \cdot \boldsymbol{\theta}_j\right) \psi_{j}(\boldsymbol{\ell}).
\end{eqnarray}
The angular correlation will be given by
\begin{eqnarray}
	\label{eq:clpsi}
	\nonumber C_{\ell, \mbox{\scriptsize total}}^{\psi\psi} &=& \left\langle \psi_{\mbox{\scriptsize total}}(\boldsymbol{\ell}) \psi^{*}_{\mbox{\scriptsize total}}(\boldsymbol{\ell})\right\rangle, \\
	\nonumber & = & \sum_{j} C^{\psi\psi}_{\ell, j} + \sum_{j \neq k} \exp\left( -i \boldsymbol{\ell} \cdot\Delta\boldsymbol{\theta}_{jk}\right) \left\langle \psi_{j}(\boldsymbol{\ell})\psi^{*}_{k}(\boldsymbol{\ell})\right\rangle, \\
	C_{\ell, \mbox{\scriptsize total}}^{\psi\psi} &=& \sum_{j} C^{\psi\psi}_{\ell, j} + \mathcal{D},
\end{eqnarray}
where
\begin{equation}
	\mathcal{D} \equiv \sum_{j \neq k} \exp\left( -i \boldsymbol{\ell} \cdot\Delta\boldsymbol{\theta}_{jk}\right) \left\langle \psi_{j}(\boldsymbol{\ell})\psi^{*}_{k}(\boldsymbol{\ell})\right\rangle
\end{equation}
is the small-scale correction due to the misalignment and $\Delta\boldsymbol{\theta}_{jk} \equiv \boldsymbol{\theta}_{j} - \boldsymbol{\theta}_{k}$. Exploiting the symmetry of the system,
\begin{equation}
	\left\langle \psi_{j}(\boldsymbol{\ell})\psi^{*}_{k}(\boldsymbol{\ell}) \right\rangle = \left\langle \psi_{k}(\boldsymbol{\ell})\psi^{*}_{j}(\boldsymbol{\ell}) \right\rangle.
\end{equation}
Therefore,
\begin{eqnarray}
	\label{eq:corrections}
	\mathcal{D} & = & 2 \sum_{j < k}  \cos(\boldsymbol{\ell} \cdot \Delta\boldsymbol{\theta}_{jk}) \left\langle \psi_{j}(\boldsymbol{\ell})\psi^{*}_{k}(\boldsymbol{\ell})\right\rangle.
\end{eqnarray}
By performing an average over the angle $\boldsymbol{\ell} \cdot \Delta\boldsymbol{\theta}_{jk}$ in Eq.~(\ref{eq:corrections}) and exploiting the relation
\begin{equation}
	\frac{1}{2\pi}\int_{0}^{2\pi}\mbox{d}\phi\ \cos\left(x\cos\phi\right) = J_{0}(x),
\end{equation}
where $J_n(x)$ is the Bessel function of the first kind.  Hence, the angular power spectrum of the system is given by
\begin{equation}
	C_{\ell, \mbox{\scriptsize total}}^{\psi\psi} = \sum_{j} C^{\psi\psi}_{\ell, j} +  2 \sum_{j < k} J_0(\ell \Delta\theta_{jk}) \left\langle \psi_{j}(\boldsymbol{\ell})\psi^{*}_{k}(\boldsymbol{\ell})\right\rangle.
\end{equation}

\end{document}